\documentclass[12pt]{iopart}
\usepackage{url}

\usepackage{graphicx} 
\usepackage{tabularx}
\usepackage{iopams}
\expandafter\let\csname equation*\endcsname\relax
\expandafter\let\csname endequation*\endcsname\relax
\usepackage{amsmath}
\graphicspath{ {./images/} }
\usepackage{caption}
\usepackage{float} 
\usepackage{xcolor}
\usepackage{natbib}
\bibpunct[, ]{(}{)}{,}{a}{}{,}%
\usepackage{multirow}
\usepackage{lscape}
\usepackage{lineno}

\begin{document}

\title[Alpha Irradiation Platform]{The Development of a Preclinical Alpha Irradiation Platform with Versatile Control of Dose, Dose Rate, and Spatiotemporal Irradiation Patterns}

\author{Harsh Arya\textsuperscript{1}, Joshua Rajan\textsuperscript{1}, Varghese Chirayath\textsuperscript{1}, Mainul Arbar\textsuperscript{1}, Awat Lotfihagh\textsuperscript{1}, Sitmukhambetov Satzhan\textsuperscript{1}, Yan Chang\textsuperscript{2}, Alex Weiss\textsuperscript{1}, Zui Pan\textsuperscript{2}, Yujie Chi\textsuperscript{1,*}}

\address{\textsuperscript{1}Department of Physics, The University of Texas at Arlington, Arlington, TX.\\
\textsuperscript{2}Department of Graduating Nursing, The University of Texas at Arlington, Arlington, TX.}
\ead{yujie.chi@uta.edu}
\vspace{10pt}

\begin{abstract}
\textit{Objectives.} The objective of this study is to develop and validate a fully integrated, vacuum-based alpha irradiation platform to support preclinical radiobiological studies. We aim to demonstrate precise, independent control over key physical parameters, such as particle entry energy, fluence rate, and spatiotemporal radiation patterns, which are critical to the radiobiological mechanisms underlying targeted alpha therapies and low-dose risk assessments. \textit{Approach.} A vacuum-based irradiation system incorporating a radioactive alpha source was designed, fabricated, and validated. The platform provides independent modulation of key irradiation parameters: (i) temporal exposure patterns via a programmable gate valve; (ii) fluence rate across two orders of magnitude by varying the source-to-aperture distance (57 to 381 mm); (iii) incident alpha-particle energy ($0$ to $4.6$ MeV) using adjustable aperture-to-sample absorption layers; and (iv) spatial exposure distributions using a programmable 3D motion stage. Temporal precision was assessed via synchronized audio-electronic recordings of valve actuation. Fluence rates and average energies were validated using CR-39 detectors compared with Monte Carlo (MC) simulations. Spatial precision was verified using programmed continuous and discrete trajectories. \textit{Main results.} Validation experiments demonstrated high system fidelity. Measured irradiation durations deviated from programmed values by less than $0.3$ s. Measured and computed fluence rates showed excellent agreement, with differences within $3\%$. For energy validation, experimentally determined CR-39 track diameters matched MC model predictions within one standard deviation. Additionally, recorded spatial irradiation patterns and dimensions aligned well with programmed motion trajectories. \textit{Significance.} We have successfully developed and validated a versatile vacuum-based alpha-irradiation platform that overcomes the energy-degradation constraints of traditional gas-filled systems. By providing precise, multi-parametric control over alpha-particle delivery, this system enables systematic investigation into the influence of energy, dose rate, and spatiotemporal patterns on radiobiological responses. This platform is poised to optimize targeted alpha therapies and refine radiation protection frameworks for low-dose alpha exposure.
\end{abstract}

\section{Introduction}
Alpha particles emitted from radionuclides typically have energies of several mega electron volts (MeV), which are associated with high linear energy transfer (LET), short penetration ranges, and the formation of densely ionizing tracks when they traverse biological tissues. These tracks produce clusters of complex and often irreparable DNA damage, which form the basis of the long-standing view that alpha-particle induced cytotoxicity arises primarily from direct, targeted damage to the cell nucleus \citep{pouget2021revisiting}. However, growing evidence from both clinical and preclinical studies indicates that alpha irradiation can also induce non-targeted, or bystander, effects in cells located beyond the physical range of the emitted particles \citep{boyd2006radiation, widel2017radionuclides}. Mechanistic studies suggest that these effects originate from damage to non-nuclear structures such as cytoplasm, plasma membrane, and mitochondria, which in turn activate intercellular signaling cascades capable of propagating DNA damage responses to neighboring cells \citep{pouget2021revisiting, morgan2003non, azzam2002oxidative}. These non-targeted effects may contribute to as much as 30\% of cell death observed in targeted alpha therapy (TAT) \citep{pouget2021revisiting, makvandi2018alpha, elgqvist2014potential, parker2018targeted} and are thought to underlie the phenomenon of hypersensitivity to low-dose alpha irradiation \citep{heuskin2014low, nagasawa1999unexpected}. 

More critically, recent studies have also shown that the magnitude and characteristics of these non-targeted cellular effects are radiation condition and cellular microenvironment dependent, including factors such as LET, absorbed dose distribution, fluence rate, and oxygen levels \citep{pouget2021revisiting}. This contrasts with the traditional DNA-centered paradigm, which assumes that alpha particle induced DNA damage is largely independent of radiation dose rate and cellular oxygenation. These findings underscore the need for an integrated view of alpha radiobiology to improve the understanding and prediction of radiobiological responses \citep{tronchin2022dosimetry}. Such an approach will advance the design and optimization of alpha-based therapies, like source positioning in diffusing alpha-emitter radiation therapy (DART) \citep{poty2018alpha, heger2023alpha} and dose prescription in both DART and TAT \citep{guerra2020targeted, kim2024comprehensive}. It will also enable a better understanding of health risks associated with low-dose alpha exposure \citep{herrera2022low, pouget2021revisiting}.

To meet this research need, experimental systems capable of precisely controlling exposure conditions are required. A critical challenge for building such systems arises from the high LET and limited penetration range of alpha particles with energies of several MeV. In this regard, a vacuum-based particle delivery system is particularly well suited, as it is energy degradation free and has sufficient space for integrating additional control components. However, early attempts to develop such systems were unsuccessful \citep{barendsen1960effects}, leading to the adoption of helium and other gas-filled designs in subsequent radioactive source-based alpha irradiation platforms \citep{barendsen1960effects, roos1989design, inkret1990radiobiology, ishigure1991device, neti2004multi, soyland2000new, wang2005bystander, beaton2011development, szeflinski2020radiobiological, nikitaki2021construction, esposito2006244cm, esposito2009alpha, babu2013dosimetry, lee2016practical, nawrocki2018design, thompson2019tracking}. These gas-filled designs reduced energy loss rate compared with exposure in air. However, accumulated energy loss during transport still makes it limited in radiation parameter controls.

In this study, we revisit the concept of vacuum-based alpha particle delivery using modern materials, precise engineering, and advanced control systems. We have developed a fully integrated vacuum-based alpha irradiation platform that enables precise and versatile control of particle entry energy at the sample plane, as well as fluence rate, and both spatial and temporal radiation patterns. The system was validated through comprehensive physical characterization and biological experiments. In the following, we describe the design principles, fabrication, construction, and physical validation design of the platform in the Method section, following by the corresponding results in the Result section, which ends with a discussion and conclusion.

\section{Methods}

\subsection{Overall Working Principle}

\begin{figure}[H]
    \centering
    \includegraphics[width=\textwidth]{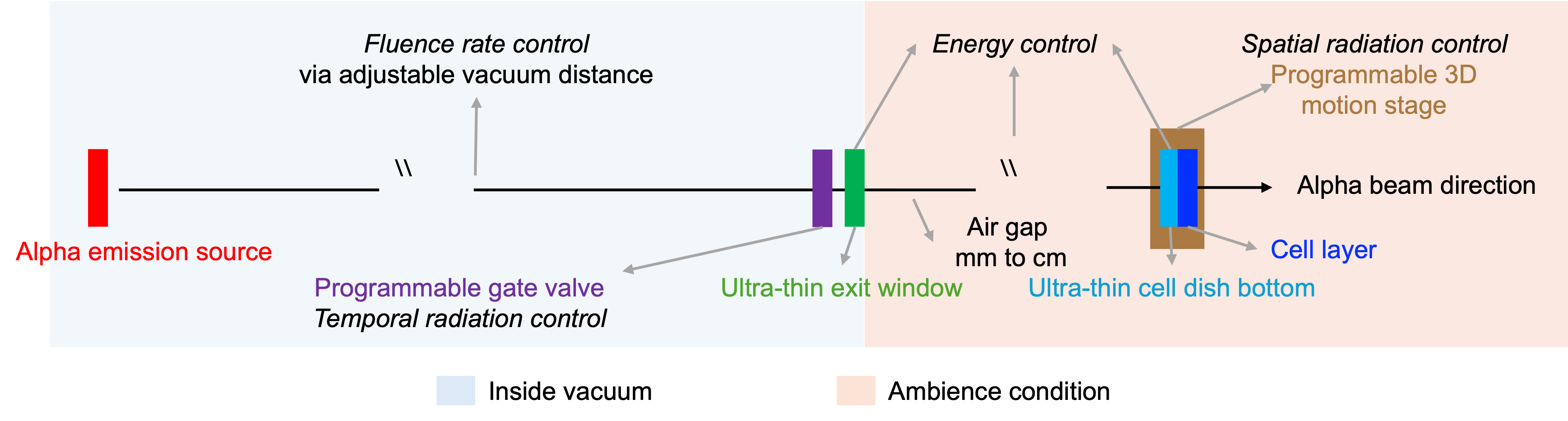}
    \caption{The overall design principle of the alpha irradiation platform. The system features vacuum-based fluence rate control, programmable gate valve for temporal radiation modulation, exit window--air gap--cell dish bottom stack for energy control, and a programmable 3D motion stage for precise spatial radiation delivery.}
    \label{fig: overall principle}
\end{figure}
The overall design of the irradiation system is illustrated in Fig. \ref{fig: overall principle}. The entire system is divided into two regions: a vacuum system that houses an alpha-emitting source and an ambient space that accommodates biological samples. To independently regulate alpha fluence rate, its source position  within the vacuum chamber is adjustable, thus to change the source-to-aperture distance. Temporal radiation pattern is controlled by a programmable gate valve that opens and closes according to predefined timing sequences and precise spatial control of the dose distribution is achieved by translating the biological samples using a programmable 3D motion stage. Final alpha particle energy at the sample surface is then controlled by the several material layers it needs to pass through in ambiance, such as vacuum exit window, air gap, and sample holder. Collectively, the system provides precise and independent control of particle energy, fluence rate, and temporal and spatial radiation patterns for in-vitro radiobiological studies.
\subsection{Design, Material, and Fabrication} \label{sec:design}
\begin{figure}[H]
    \centering
    \includegraphics[width=\textwidth]{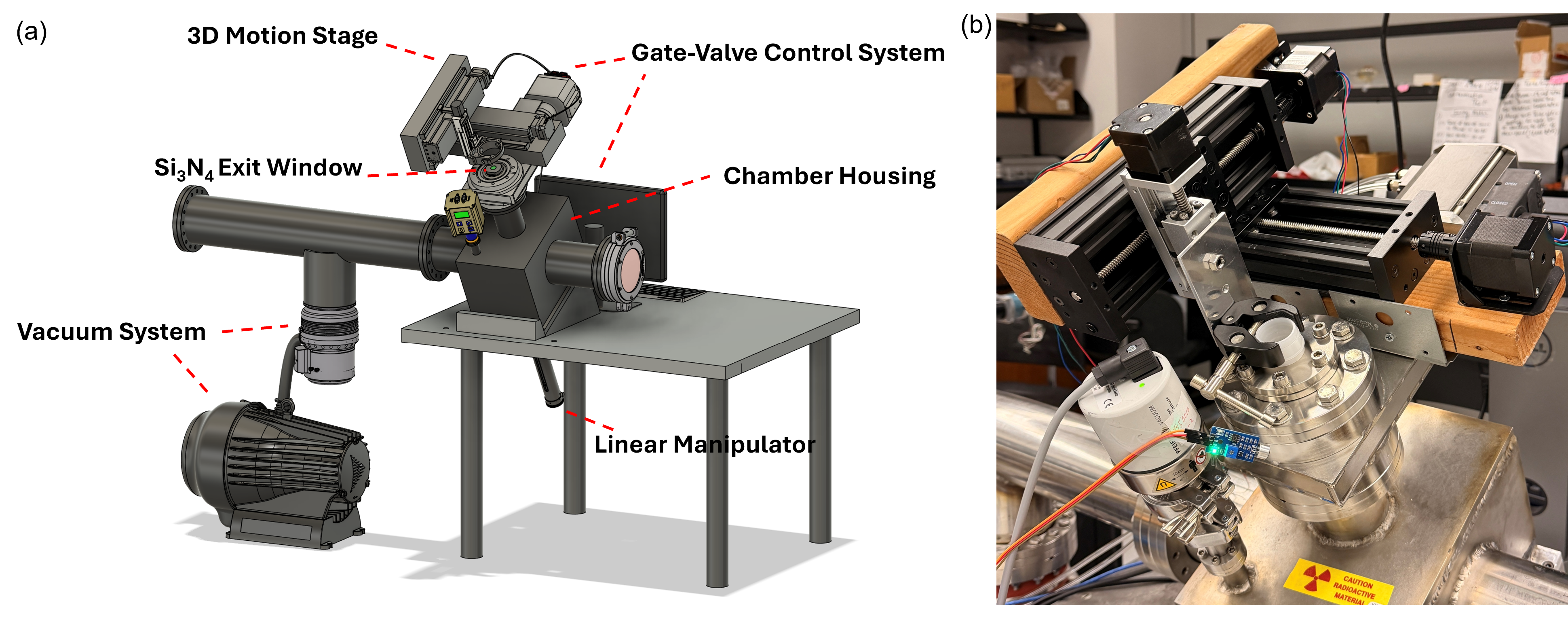}
    \caption{Schematic illustration and physical implementation of the alpha delivery system. (a) 3D design model showing major components. (b) Photograph of the fabricated system.}
    \label{fig:platform design}
\end{figure}

Based on the overall working principle, we designed and constructed a practical alpha delivery system, as illustrated in Fig.~\ref{fig:platform design}. Each component was either obtained from a vendor or fabricated in the university machine shop, and subsequently assembled to form the complete system. The remainder of this subsection describes the material specifications and fabrication details for each component. In the next subsection, we have the assembly process and system testing.

\begin{figure}[H]
    \centering
    \includegraphics[width=0.8\textwidth]{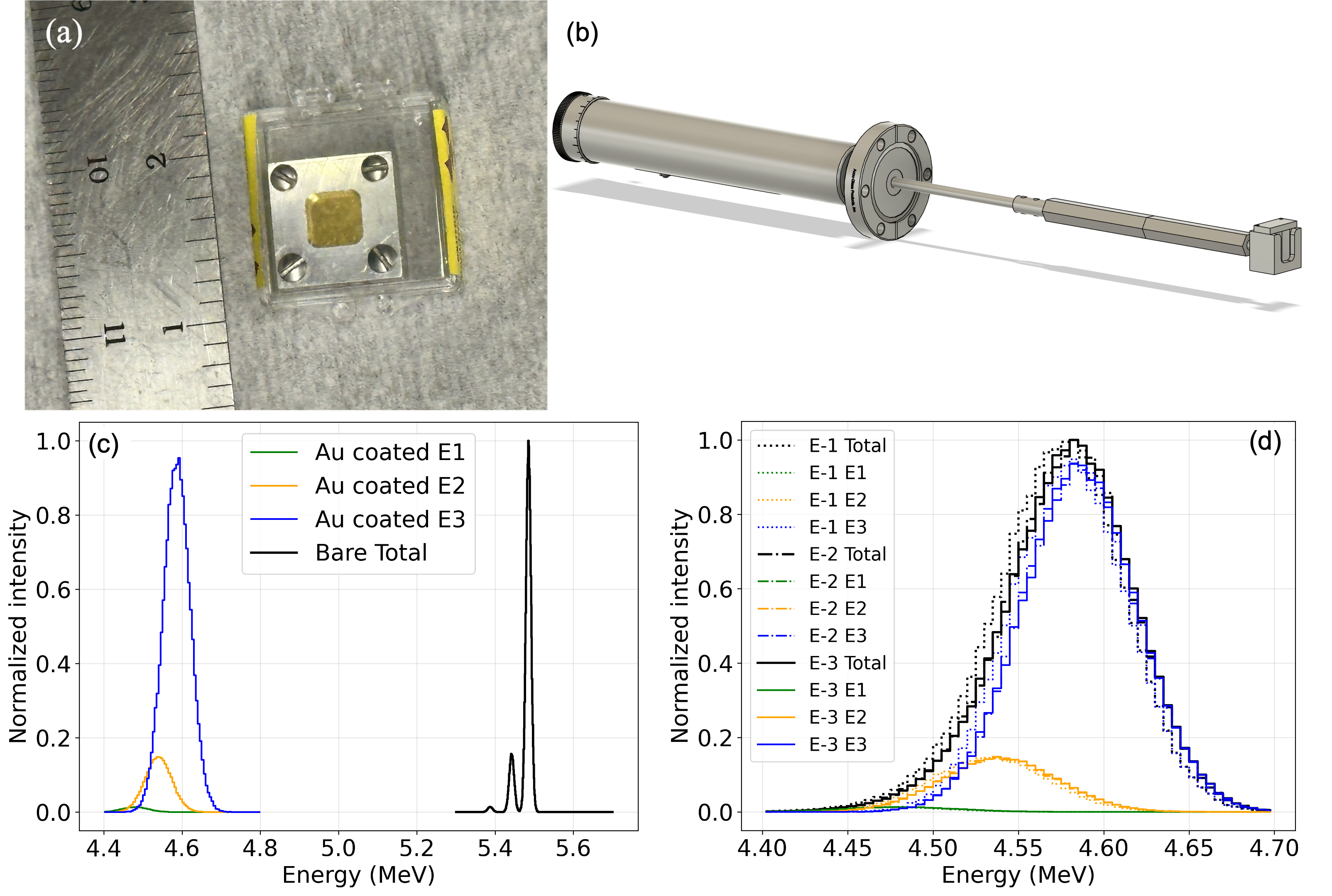}
    \caption{(a) Am-241 source placed inside a plexiglass box, with a ruler shown for scale.(b) Source mounting system installed inside the vacuum chamber.(c) Geant4-simulated alpha energy spectra before (``bare'') and after (``Au-coated'') transmission through a 2.0~$\mu$m gold layer.(d) Alpha energy spectra after transmission through a 330~mm vacuum path at pressures of $10^{-1}$, $10^{-2}$, and $10^{-3}$~hPa. Here, E1, E2, and E3 correspond to the three primary alpha emission energies of the Am-241 source: 5.388~MeV, 5.443~MeV, and 5.486~MeV, respectively.}
    \label{fig:alpha source and energy}
\end{figure}

Am-241 was chosen as the alpha-emitting source considering its long half-life (432.2 years), clean decay process dominated by alpha emission, and stable daughter nuclei (Np-237). The emitted alpha particles have energies of 5.486~MeV (85.2\%), 5.443~MeV (13.4\%), and 5.388~MeV (1.2\%), accompanied by a gamma emission with a dominant energy of 59.54~keV. A vacuum-friendly Am-241 source with a total activity of 4.8~MBq was obtained from Eckert \& Ziegler Isotope Products. The source was fabricated as a sandwich structure embedded within an aluminum holder, in which americium oxide was encapsulated between two gold foils, with the upper foil measuring 2~$\mu$m in thickness. The aluminum holder has external dimensions of 0.75~$\times$~0.75~in$^2$ with a thickness of 0.13~in, and provides an active alpha-emitting area of 0.32~$\times$~0.32~in$^2$ (Fig.~\ref{fig:alpha source and energy}a). According to vendor’s specification, energy spread out for the dominant alpha emission energy is 6.4 keV. We assumed a same energy spread-out for the other two alpha energies, and performed Geant4-based simulations \citep{agostinelli2003geant4} to obtain the alpha energies after passing through the 2.0~$\mu$m gold layer. The result is shown in Fig.~\ref{fig:alpha source and energy}(b), where the dominant energy decreases to 4.586~$\pm$~0.034~MeV. 

With it, we simulated alpha particle energy loss over a 330~mm travel distance under different vacuum pressures. As is shown in Fig.~\ref{fig:alpha source and energy}(c), energy degradation and spectrum spread out is negligible (below 0.1\%) even under 10$^{-1}$ hPa. Based on it, we selected a Pfeiffer HiScroll~12 backing pump for vacuum maintenance and connected it to a vacuum chamber fabricated from SS304 stainless steel with dimensions of 10~in~$\times$~7~in~$\times$~7~in. The chamber has a wall thickness of 0.625~in to provide adequate shielding against gamma radiation. The vacuum pressure was measured using a Pfeiffer ActiveLine PKR~251 combination gauge, interfaced with a Pfeiffer DCU~310 control and display unit mounted on the chamber’s top flange.

The Am-241 source was mounted into the vacuum chamber through a bottom flange using a three-stage mounting system (Fig.~\ref{fig:alpha source and energy}c). The source was first secured in a stainless-steel holder, which was connected to two standoff bars, itself attached to a vacuum-sealed linear manipulator (Accu-Glass Products). The manipulator provided a travel range of 0-101.6 mm with a precision of 1~mm. Together with the standoff bars, source-to-exit window distance can be adjusted between 57 mm and 381 mm. Once mounted, the source position could be visually verified through an observation window consisting of a VPG-3.8-600 Zero-Length CF viewport (Accu-Glass Products) fitted with Larson Electronic Glass (Model VP-400-F6, Part No. 111272) made from Corning 7056 borosilicate glass, providing enhanced viewing and radiation protection. 

A VAT Series~10.8 UHV gate valve was mounted on top of the vacuum chamber and interfaced with a pneumatic solenoid system controlled by a Raspberry Pi microcontroller via a relay module for temporal control. Two primary software scripts were developed: one enabling timed exposures and the other providing a shutdown response to the rupture of the exit window via an acoustic signal collected by a microphone. Upon receiving an open or close command, the Raspberry Pi output a 3.3~V digital signal to the relay, which switched a 24~V control line powering the solenoid valve. The solenoid directed compressed air (60--80~psi) to either the open or close port of the gate valve’s pneumatic actuator, thereby providing the pressure required for both opening and closing the gate valve.

Mounted above the gate valve was the exit window. Based on the estimated vacuum level and the requirement to minimize alpha energy loss, a silicon nitride membrane with a thickness on the order of a few hundred nanometers, supported by a 200 µm-thick silicon frame (Ted Pella, Inc.), was selected. The membrane assembly was mounted onto the gate valve using a DN CF-40 flange (Silson Ltd.).

An in-house built 3D motion stage was mounted above the exit window for biological sample positioning. The stage consisted of motorized linear rails for the $X$-, $Y$-, and $Z$-axes, with the $Z$-axis oriented perpendicular to the exit window. Linear translation along the $X$ and $Y$ axes was driven by NEMA~17 stepper motors, with the $Z$-axis by a compact NEMA~14 stepper motor. The motors were controlled by the Rasberry Pi for stage trajectories, dwell times, and movement intervals.
The stage provided a total travel range of 120 mm in both $X$ and $Y$ directions, and a smaller stroke length of 50 mm for the $Z$ direction.
The stage platform supported 3D-printed cell-culture dishes or Chemplex~1530-SE sample cups bottomed with micrometer-thick Mylar film (Chemplex~3012), with the film secured to the dishes using compression rings.

\subsection{System Performance Testing}
Following machining, all stainless-steel components were cleaned according to standard vacuum-preparation procedures to remove particulates and organic residues. After cleaning, the components were assembled and gone through leak testing using a helium mass spectrometer with the pump operating.

The fully assembled chamber achieved a base pressure of $2.0 \times 10^{-3}$~hPa or better within 30~minutes of pumping, which remained stable over 24-hour periods with negligible drift. We then performed performance testing on temporal and spatial control, fluence rate, and energy-tuning capabilities of the system.

For all performance tests, CR-39 based alpha-particle track detection was employed. CR-39 is a polymeric solid-state nuclear track detector (SSNTD) widely used in charged-particle detection and dosimetry \citep{fews1982high, rana2018cr, khan2023efficient}. When an alpha particle traverses the detector material, it produces a latent damage trail along its path. These latent tracks are subsequently enlarged and revealed through chemical etching, producing micron-scale pits that can be visualized and counted under an optical microscope. By identifying and counting individual tracks, the number of alpha particles reaching the detector surface can be determined. Furthermore, because the etching rate along the damage trail differs from that of the undamaged bulk material, the resulting pit geometry (e.g., radius or cone angle) can be correlated with the particle’s incident energy, enabling the measurement of alpha energy.

In this study, CR-39 detectors named TASTRAK$^\text{TM}$ (Track Analysis Systems Ltd., UK) were used, each measuring 2.5~cm~$\times$~2.5~cm~$\times$~1.5~mm. Following irradiation, the detector was etched in 6.25~M NaOH at 98$^\circ$C for 1~h, rinsed with deionized (DI) water, immersed in 5\% diluted acetic acid for 5~min, rinsed again with deionized (DI) water, and gently blotted dry. The etched CR-39 samples were examined using a Leica optical microscope at 5$\times$ magnification for track density quantification and 20$\times$ for energy examination. To reduce background noise and extend detector shelf life, CR-39 sheets were stored at $-20\,^\circ$C prior to use.

\subsubsection{Temporal Control}\hfill\\
As described in Section~\ref{sec:design}, the Raspberry Pi controlled pneumatic gate valve undergoes a sequence of pneumatic and mechanical reaction processes that determine its temporal behavior during radiation delivery. Delays in adjacent operations may introduce discrepancies between the programmed exposure duration and the actual exposure duration (Mode 1), and may also influence the responsiveness of the system under emergency shutoff conditions (Mode 2).

To test its impact in Mode 1, we programmed the valve to open for predefined exposure durations (\(T_{\text{set}}\)) of 5, 10, and 15~s before closing. A Geiger–Müller (GM) counter was positioned exactly above the exit window, and a 240 fps camera was used to record the GM response, providing an audible indication of radiation transmission. We then defined the interval between the start and ending of the audio signal of the GM as real exposure time (\(T_{\text{real}}\)). Meanwhile, sound of air entering the actuator or venting during closure, were used to identify the open and close reaction intervals. Each condition was measured three times. 

For the emergency shutdown test (Mode 2), a sound pulse was manually generated to trigger the acoustic sensor, which sent a digital signal to the Raspberry~Pi controller to close the gate valve. Video and GM counter recordings were again used to determine the total delay between sound detection, control signal transmission, and termination of alpha exposure.

\subsubsection{Fluence Rate Testing}\hfill\\
The fluence rate of alpha particles transmitted through the exit aperture was first computed theoretically and then verified experimentally. The theoretical estimation was performed using a Monte Carlo (MC) integration approach that considered only the geometric relationship between a uniformly emitting planar source and the exit aperture. Given the source activity $A$, active source area ($L \times W$), aperture area ($L_a \times W_a$), and the separation distance $d$ between the source plane and the aperture plane, the averaged particle fluence rate $\dot{\Phi}$ (particles·s$^{-1}$·mm$^{-2}$) reaching the aperture can be expressed as:
\begin{align}
\dot{\Phi}(d) =
\frac{A}{4\pi L W L_a W_a}
& \int_{-L/2}^{L/2} \! dx
  \int_{-W/2}^{W/2} \! dy  \notag \\[2pt]
& \times
  \int_{-L_a/2}^{L_a/2} \! dx_m
  \int_{-W_a/2}^{W_a/2} \! dy_m
  \frac{d}{\left[(x - x_a)^2 + (y - y_a)^2 + d^2\right]^{3/2}} .
\label{eq1}
\end{align}

Here, a Cartesian coordinate system is defined based on the source, with the source center as the origin, and the $x-y$ axes aligned with the source edges. $(x, y)$ and $(x_m, y_m)$ are point coordinates on the source and the aperture, respectively. Further, for a sample-plane placed at a distance $g$ downstream the exit aperture, geometric penumbra widens the effective radiation area. Assuming straight-line particle trajectories, the maximum area dimensions are obtained by considering vertex-to-vertex trajectories connecting the rectangular source and aperture, which result in an effective sample irradiation area as 
\begin{equation}
S_{\mathrm{eff}}
=
\left( L_a + \frac{g}{d}\left( L_a + L \right) \right)
\left( W_a + \frac{g}{d}\left( W_a + W \right) \right) ,
\label{eq2}
\end{equation}
which gives the effective average dose rate at the sample as
\begin{align}
\dot{\Phi}_{\mathrm{eff}}(g,d)
= \frac{{L_a W_a}}{S_\mathrm{eff}}\,\dot{\Phi}(d) .
\label{eq3}
\end{align}

For the present setup, the source activity was $A = 4.8$~MBq, with an active emitting area of $L = W = 8.128$~mm. The exit window contained a 2~$\times$~2~mm$^2$ aperture that defined the effective aperture area ($L_a$ and $W_a$). The source-to-aperture distance $d$ was adjustable between 57~mm and 381~mm. The aperture-to-sample distance $g$ was adjustable between 0.5~mm and 20.5~mm. Using these parameters, numerical integration was performed to obtain the fluence rate $\dot{\Phi}$ as a function of distance $d$. To show the impact of $g$ over both high and low fluence rate regions, we selected two reference $d$ points as $d_1=61$~mm and $d_2=279$~mm to compute $\dot{\Phi}_{\mathrm{eff}}$ as a function of $g$.

To validate the theoretical calculation at the aperture plane, we experimentally measured the alpha-particle fluence rates using CR-39 based track detection. The source-to-exit window distance $d$ was set to two positions, 279~mm and 381~mm, while the exit window was maintained as a 2~$\times$~2~mm$^2$ aperture with a membrane thickness of 100~nm. The CR-39 detector was placed directly on top of the exit window and exposed to alpha radiation for 15~s, with exposure timing precisely controlled by the programmable gate valve. Following irradiation, the detector was etched and track density quantified according to the standard protocol described above.

\subsubsection{Alpha Energy Testing}\hfill\\
To evaluate the energy modulation effect of the combined air and Mylar film layers, we performed CR-39-based energy measurements for the following three setups: 1) 1~mm air gap with 0~$\mu$m Mylar layer; 2) 1~mm air gap with 2.5~$\mu$m Mylar layer; and 3) 1~mm air gap with 7.5~$\mu$m Mylar layer. For all measurements, the source-to-exit window distance was fixed at 279 mm, the exit window was kept to be a 2~$\times$~2~mm$^2$ aperture with a membrane thickness of 100~nm, and the exposure time was set to 15~s. Two CR-39 based independent measurements were performed for each condition. After irradiation, the CR-39 detectors were chemically etched and the resulting track diameters (\(D\)) were measured. We then verified the measured track diameter for each condition via the following computational procedure.

\begin{figure}[H]
    \centering
    \includegraphics[width=1\textwidth]{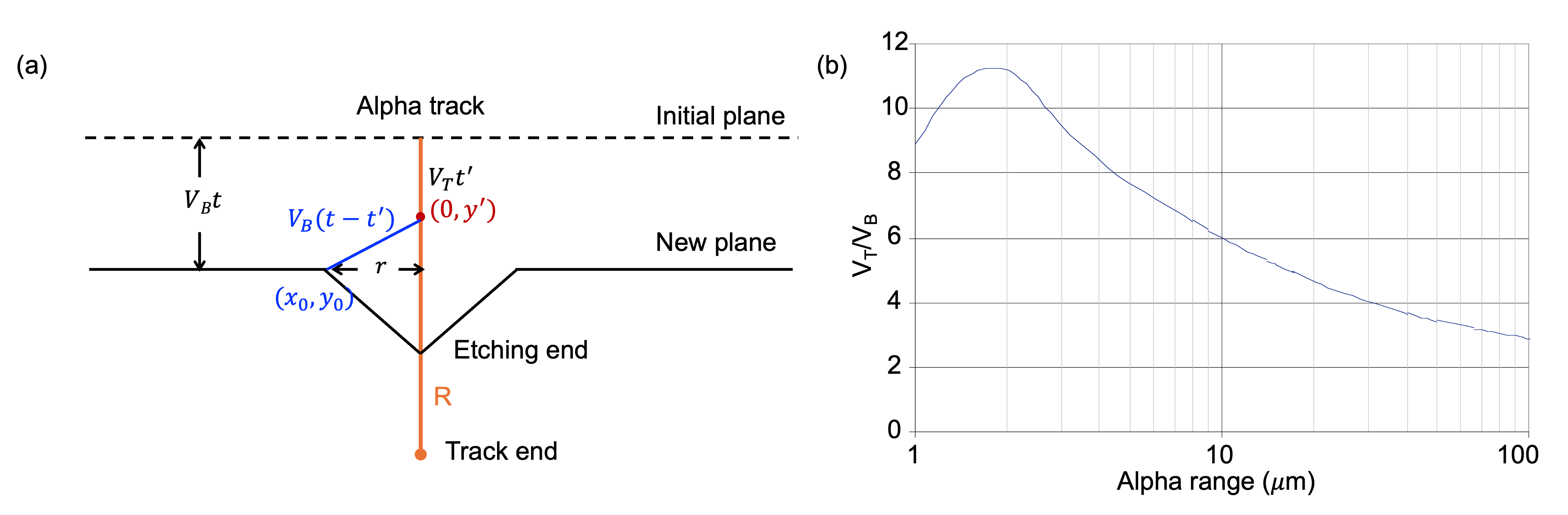}
    \caption{Principle of alpha-particle energy determination from CR-39 track measurements. (a) Geometry of etch pit formation for alpha particles with different energies corresponding to ranges \(R_1\) and \(R_2\). (b) Relationship between the track-to-bulk etch rate ratio (\(V_T/V_B\)) and the alpha-particle range in CR-39, as provided by Track Analysis Systems Ltd., UK 
    (https://www.tasl.co.uk/downloads/tastrak-alpha-specifications.pdf).}
    \label{fig:alpha_energy_measurement}
\end{figure}

We first correlate the measured track diameter to the chemical etching kinetics. As illustrated in Fig.~\ref{fig:alpha_energy_measurement}(a), the CR-39 bulk material etches uniformly at a rate \(V_B\), whereas the latent alpha track etches at a higher track rate \(V_T\). Consider an alpha particle entering the detector normal to the surface and leaving a latent track of total length \(R\). After an etching duration \(t\), bulk etching lowers the detector surface by \(V_B t\). Because \(V_T \gg V_B\), a conical etch pit of radius \(r\) forms at the new surface. To derive \(r\), we follow the standard geometric construction. At an intermediate etching time \(t' < t\), the track has been etched back a distance \(V_T t'\), reaching the point \((0, y')\) along the track direction. Around this point, bulk etching over the remaining time interval \((t - t')\) removes material to a depth of \(V_B(t - t')\). For a correct value of \(t'\), the sphere of bulk-etched material intersects the final etched surface at coordinates \((x_0, y_0)\), with \(|x_0| = r\). Allowing \(V_T\) to vary with the remaining alpha range, the pit radius and depth satisfy the following equations \citep{fleischer1969nuclear}:
\begin{equation}
\begin{aligned}
x_0 &= V_B \left(t - \int_{0}^{y'} \frac{dy}{V_T(y)}\right)
      \left(1 - \frac{V_B^2}{V_T(y')^2}\right)^{1/2}, \\
y_0 &= y' + \frac{V_B^2}{V_T(y')}
      \left(t - \int_{0}^{y'} \frac{dy}{V_T(y)}\right).
\end{aligned}
\label{eq4}
\end{equation}

To solve Eq.~(\ref{eq4}), both the bulk etch rate \(V_B\) and the ratio \(V_B/V_T(y)\) as a function of residual track depth are required. According to the detector vendor (Track Analysis Systems Ltd.), the ratio \(V_B/V_T\) is empirically correlated with the residual alpha range \(R\) in CR-39, as shown in Fig.~\ref{fig:alpha_energy_measurement}(b), and the bulk etch rate is specified as around \(V_B = 10~\mu\text{m/h}\). We then performed SRIM-based MC simulations \citep{ziegler2010srim} to obtain the relationship between the residual range \(R\) and the incident alpha-particle energy \(E\). Simulations were conducted from 0.1 to 5.0~MeV in 0.1~MeV increments, producing 50 discrete energy points with \(10^4\) particles per run.

Geant4-based simulations were subsequently used to determine the incident alpha energy reaching the CR-39 detector under each experimental condition. By combining these incident-energy estimates with the vendor-provided \(V_B/V_T(R)\) correlation and the SRIM-derived \(E\)--\(R\) mapping, we obtained \(V_T(y)\) for each condition. We then estimated the etch-pit diameters by numerically solving Eq. \ref{eq4}. Considering \(V_B\) can be detector dependent, we first calibrated \(V_B\) under condition 1) and then applied it to compute track diameters for conditions 2) and 3), which were compared with measured CR-39 track diameters. Geant4-simulated incident energy uncertainty was further used to predict track diameter uncertainty via MC based uncertainty propagation with 5,000 samples per condition.

\subsubsection{Spatial Pattern Control}\hfill\\
To evaluate the spatial radiation control of the system, two irradiation configurations were implemented: a continuous rectangular exposure and a discrete dual-spot exposure. In both tests, the source-to-exit-window distance was set to 279 mm, and CR-39 detectors were placed on the motion-stage holder 10 mm above the aperture to record the resulting alpha-track distributions.

For the continuous exposure test, the stage was programmed to follow a closed rectangular path at a constant velocity of $0.25$ mm/s over a total duration of $20$ minutes. Starting from the detector center, the stage was translated sequentially: $4$ mm downward, $5$ mm to the left, $8$ mm upward, $5$ mm to the right, and finally $4$ mm downward to return to the origin. The gate valve remained fully open throughout the sequence to ensure uninterrupted alpha-particle delivery along the trajectory.

To verify the synchronized actuation of the gate valve with stage movement and assess spatial reproducibility, a discrete irradiation test was performed. A $2 \times 2$ mm$^2$ irradiation field was initially positioned at the lower-left corner of the CR-39 detector for a $15$ s exposure. Following this, the stage was translated diagonally to the top-right corner, where a second $15$ s exposure was conducted. The gate valve was programmed to close during stage translation, ensuring two distinct, separate exposure events.

\section{Results} \label{Results}
\subsection{Temporal Control Testing Results}
\begin{figure}[H]
    \centering
    \includegraphics[width=\textwidth]{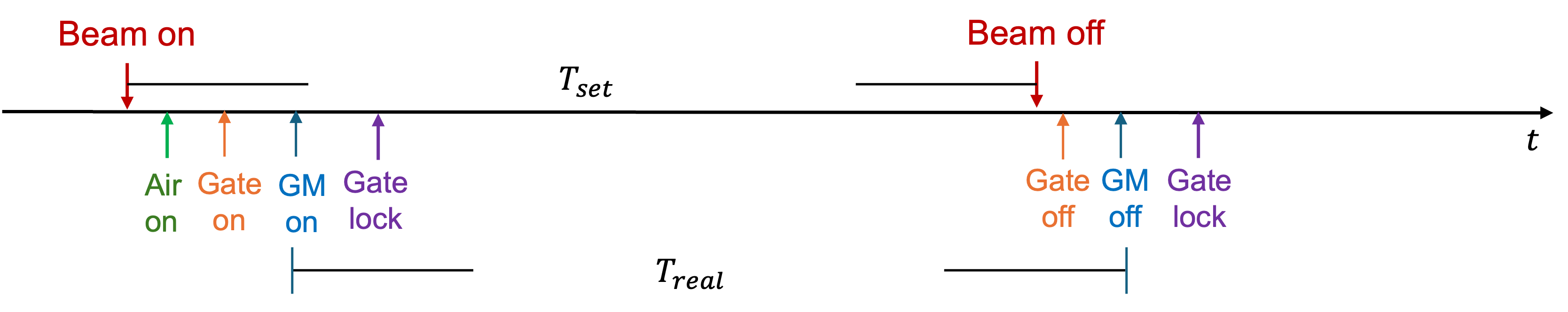}
    \caption{Time stamps for the programmed irradiation via gate valve control.}
    \label{fig:sequence}
\end{figure}

From the audio recordings, the time stamps associated with each stage of the valve actuation sequence were extracted, as illustrated in Fig.~\ref{fig:sequence}. After the program issued the beam-on command, air pressurization began (“air on”), followed by the onset of valve opening (“gate on”), the appearance of GM counts (“GM on”), and finally the fully open lock-in click (“gate lock”). After a programmed interval of \(T_\text{set}\), the controller issued the beam-off command, after which the valve began to close (“gate off”), the GM signal ceased (“GM off”), and the fully closed lock-in click was recorded (“gate lock”).

Based on the extracted time stamps, the interval between ``air on'' and ``gate off'' closely matched the programmed exposure duration \(T_\text{set}\), with measured values of \(4.98\pm0.01\)~s, \(9.99\pm0.01\)~s, and \(14.98\pm0.01\)~s for the 5, 10, and 15~s settings, respectively. However, due to a finite delay between ``air on'' and the onset of mechanical valve opening (``gate on'') during beam-on transitions, the effective radiation exposure durations (\(T_\text{real}\)) were slightly shorter than \(T_\text{set}\). The measured \(T_\text{real}\) values were \(4.73\pm0.02\)~s, \(9.71\pm0.05\)~s, and \(14.72\pm0.02\)~s, respectively.


As for the emergency response mode, upon detection of the trigger sound, the Raspberry~Pi controller initiated valve closure within approximately 0.02~s, and alpha transmission was fully terminated within 0.29~s of signal onset. These measurements confirm that the automatic emergency protocol reliably interrupts irradiation.


\subsection{Fluence Rate}
\begin{figure}[H]
    \centering
    \includegraphics[width=1\textwidth]{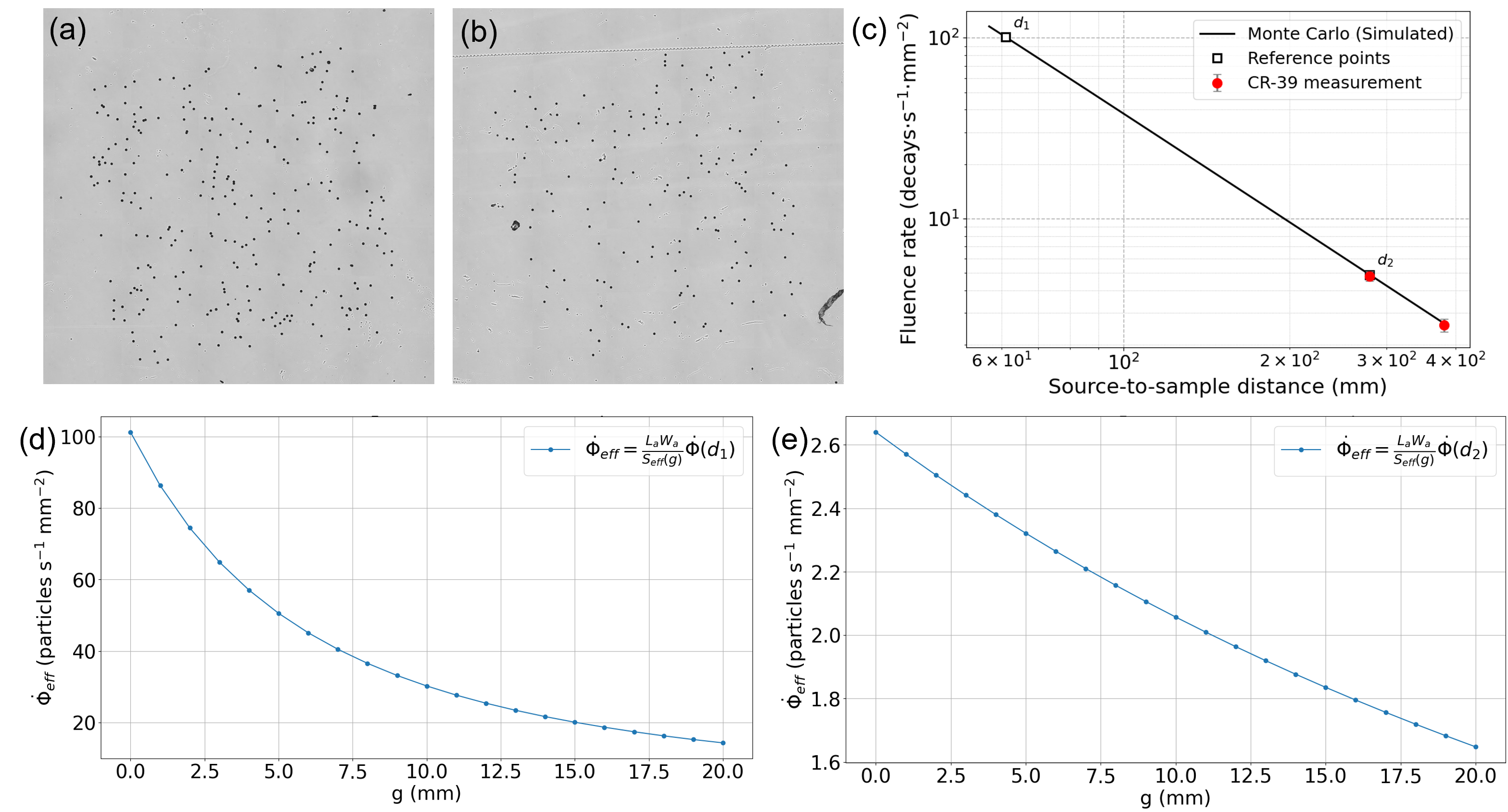}
   \caption{Alpha-particle track density observed under 5$\times$ magnification at source-to-detector distances of (a) 279 mm and (b) 381 mm, each with a 15~s exposure.(c) MC integration results showing the dependence of alpha-particle fluence rate on source-to-aperture distance, along with the two experimentally measured fluence rates. The reference distances $d_1 = 61$~mm and $d_2 = 279$~mm used for downstream fluence calculations are explicitly indicated.(d,e) Effective average dose rate at the sample plane as a function of aperture-to-sample distance $g$, computed using Eq.~(3) for $d=d_1$ and $d=d_2$, respectively.}

    \label{fig:fluence_rate_test}
\end{figure}

The theoretical fluence rate $\dot{\Phi}(d)$ defined in Eq.~(\ref{eq1}) was numerically evaluated and is shown in Fig.~\ref{fig:fluence_rate_test}(c). As the source-to-aperture distance decreases from 381~mm to 57~mm, $\dot{\Phi}(d)$ increases from 2.6 to  115.7~particles·s$^{-1}$·mm$^{-2}$. 

The raw CR-39 track counts acquired at source-to-aperture distances of 279~mm and 381~mm under 5$\times$ magnification are shown in Fig.~\ref{fig:fluence_rate_test}(a) and (b), yielding 283 and 151 tracks, respectively. Using the measured exposure duration \(T_{\text{real}} = 14.72\)~s, the corresponding measured fluence rates were 4.80 and 2.56~particles·s$^{-1}$·mm$^{-2}$. These values agree closely with MC predictions of 4.90 and 2.64~particles·s$^{-1}$·mm$^{-2}$, demonstrating excellent consistency between experiment and simulation.

The fluence rate $\dot{\phi}$ as a function of the aperture-to-sample distance $g$ at the two reference points is shown in Fig.~\ref{fig:fluence_rate_test}(d) and (e), respectively. At a short source-to-aperture distance ($61$ mm), the fluence rate exhibits high sensitivity to variations in $g$ where an increase from $0$ to $2.5$ mm in $g$ results in a sharp reduction to approximately $70\%$ of the initial fluence rate. Conversely, at a larger source-to-aperture distance ($279$ mm), the fluence rate exhibits a more gradual, approximately linear decline of roughly $60\%$ over a $20$ mm increase in $g$.

\subsection{Energy Control}

\begin{figure}[H]
    \centering
    \includegraphics[width=1.0\textwidth]{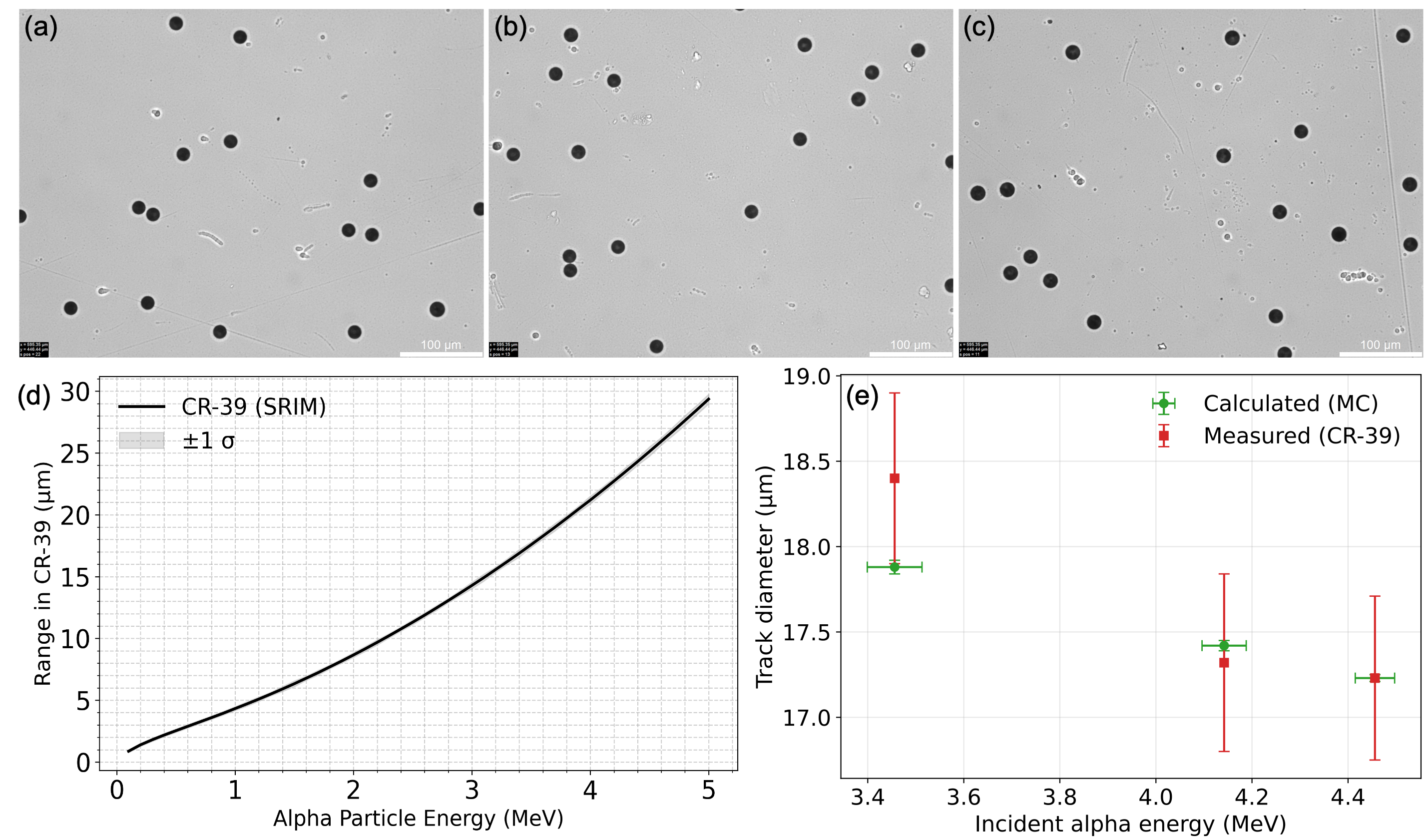}
    \caption{Representative alpha-particle tracks observed under 20$\times$ magnification with exposure conditions of (a) a 0~$\mu$m Mylar film with a 1~mm air gap, (b) a 2.5~$\mu$m Mylar film with a 1~mm air gap, and (c) a 7.5~$\mu$m Mylar film with a 1~mm air gap. (d) SRIM-simulated relationship between alpha-particle energy and range in CR-39. (e) Calculated and measured alpha track diameters as a function of alpha incident energies for the three experimental conditions.}
    \label{fig:energy_control}
\end{figure}

Representative alpha-particle tracks observed under 20$\times$ magnification for the three exposure conditions are shown in Fig.~\ref{fig:energy_control}(a)-(c). Observable differences in track diameters are shown, consistent with the expected variation in incident alpha energy. From individual track measurements, the mean~$\pm$~std.\ diameters were \(17.23\pm0.48~\mu\text{m}\), \(17.32\pm0.52~\mu\text{m}\), and \(18.40\pm0.50~\mu\text{m}\), respectively. The corresponding SRIM-derived energy–range relationship is shown in Fig.~\ref{fig:energy_control}(d).

Geant4-based simulations yielded incident alpha energies of \(4.456\pm0.041~\text{MeV}\), \(4.142\pm0.046~\text{MeV}\), and \(3.456\pm0.057~\text{MeV}\) for the three conditions. Using the SRIM-derived \(E\)–\(R\) mapping together with the vendor-provided correlation in Fig.~\ref{fig:alpha_energy_measurement}(b), the depth-dependent track-to-bulk etch-rate ratio \(V_T/V_B\) was determined. In Eq.~(\ref{eq4}), by setting \(y_0 = V_B~\mu\text{m}\), a numerical solution was performed to obtain \(y'\) and \(r = x_0\). By enforcing \(2r = 17.23~\mu\text{m}\) for condition~1, the bulk etch rate was calibrated to \(V_B = 10.73~\mu\text{m/h}\). Using this calibrated value, the predicted track diameters for the three conditions were \(17.23\pm0.02~\mu\text{m}\), \(17.42\pm0.03~\mu\text{m}\), and \(17.88\pm0.04~\mu\text{m}\), respectively. The comparison between calculated and measured diameters is shown in Fig.~\ref{fig:energy_control}(e), which shows overall good agreement between the two across all three conditions.

\subsection{Spatial Control}
\begin{figure}[H]
\centering
\includegraphics[width=1.0\textwidth]{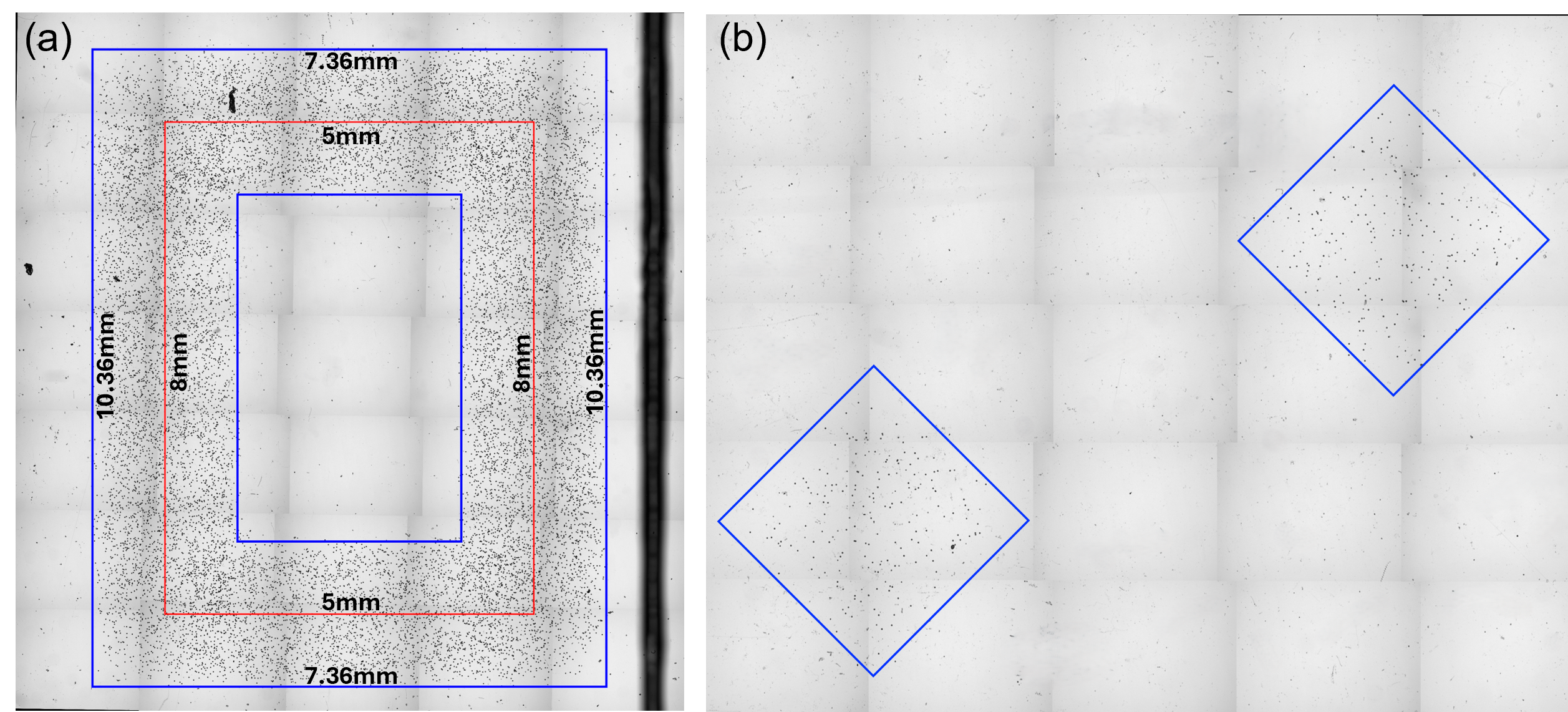}
\caption{(a) Rectangular irradiation pattern showing continuous motion control. (b) Two discrete exposure sites generated by synchronizing the motion stage position with temporal gate valve actuation.}
\label{fig:motion_stage_patterns}
\end{figure}

The continuous rectangular radiation pattern in the first test case is shown in Fig.~\ref{fig:motion_stage_patterns}(a). The processed CR-39 detector exhibits a continuous distribution of alpha tracks. Track density remains consistent along the entire path, indicating stable stage motion and uninterrupted alpha transmission during the continuous exposure. Based on Eq.~\ref{eq2}, the calculated effective radiation area for a single snapshot is $2.36 \times 2.36$ mm$^2$, resulting in a cumulative rectangular irradiation pattern of $7.36 \times 10.36$ mm$^2$. When these calculated boundaries were overlaid on the measured trajectories in Fig.~\ref{fig:motion_stage_patterns}(a), they showed excellent agreement with the spatial distribution of alpha-particle tracks.

In the second discrete irradiation test (Fig.~\ref{fig:motion_stage_patterns}(b)), two well-separated clusters of alpha tracks were observed at opposite corners of the CR-39 detector, containing approximately 240 and 236 tracks, respectively. The two clusters exhibit comparable shape and track density, indicating reliable synchronization of the programmed gate valve actuation with the programmed motion-stage movement.

\section{Discussion}

In this study, we designed and constructed a vacuum-based alpha-irradiation platform with programmable temporal and spatial radiation control, as well as independent tuning of fluence rate and incident particle energy. Its performance was evaluated through a series of validation tests. The system demonstrated accurate temporal gating, with measured irradiation durations deviating from programmed values by less than 0.3~s, and enabled fluence rate modulation spanning nearly two orders of magnitude. Energy tuning across several MeV was verified using CR-39 based measurements, and spatial irradiation control was shown to be precise, with recorded exposure patterns reproducing the programmed motion trajectories.

The platform established in this study provides a practical foundation for alpha-induced biological response studies. By allowing independent adjustment of exposure time, source-to-sample distance, incident particle energy, and spatial exposure pattern, the system is suitable for investigating a range of irradiation conditions relevant to radiobiology and radiotherapy, such as time-resolved irradiation protocols and dose-rate dependent effects.

Despite the successful operation of the platform, several limitations can be addressed in future work. As shown in Fig.~\ref{fig:energy_control}(e), while the measured and calculated track diameters agree well at an incident energy of \(4.142 \pm 0.046~\text{MeV}\), a larger discrepancy is observed at \(3.456 \pm 0.057~\text{MeV}\). We consider this difference primarily attributed to the modeling assumptions used in Eq.~(\ref{eq4}), where track-diameter calculations assume normal alpha-particle incidence. In practical irradiation conditions, however, alpha particles reach the detector with a distribution of incident angles. After traversing thicker absorption layers, increased scattering leads to a higher likelihood of oblique incidence. Such non-normal tracks are known to produce larger etch-pit diameters in CR-39 detectors compared to normal incidence \citep{fews1982high}, which likely contributes to the observed deviation between calculated and measured diameters at lower energies. Other energy measuring methods or more accurate analytical models can be used in future work for more accurate alpha energy measurements.

A critical consideration in the platform's design is the implementation of energy degradation layers. While both air and Mylar can serve as degraders, the required physical thickness for air is approximately three orders of magnitude greater than that of Mylar. For example, a 5 MeV alpha particle has a total range of approximately 35–40 mm in air at standard temperature and pressure, whereas a comparable energy loss is achieved with a Mylar foil only 20–30 $\mu$m thick. As shown in Fig. 6(d), the fluence rate decreases rapidly as this gap increases. Simultaneously, the geometric penumbra widens. This leads to a reduction in the effective dose rate and a degradation of dose uniformity across the sample plane. Thus, thin Mylar films are preferred over air gaps to minimize geometric attenuation and ensure dose uniformity.

Future developments can also expand the fluence rate tuning capability of the platform to support a broader range of experimental requirements, including ultra-low and ultra-high dose rate studies. The current tuning range of approximately two orders of magnitude is mainly constrained by the dimensions of the vacuum chamber, the travel range of the linear motion stage, and the thickness of the gate valve used in this prototype. These limitations are not intrinsic to the platform design. For instance, employing a deeper vacuum chamber together with vacuum compatible, long travel translation stages would enable substantially larger source-to-aperture distance modulation, thereby extending the accessible fluence rate range toward lower values. Conversely, replacing the existing gate valve with a thinner, fast response shutter would reduce the minimum achievable source-to-aperture distance and allow higher fluence rates. Together, these improvements provide clear pathways for enhancing the versatility of the platform while retaining its core design principles and control architecture.

\section{Conclusion}
In this work, we developed and validated a vacuum-based alpha-irradiation platform that provides programmable control over temporal, spatial, and spectral characteristics of alpha-particle exposure. It is expected to serve as a robust and flexible platform for controlled alpha radiobiological studies.
\section*{Acknowledgment}
This work is partially supported by UT Arlington under grants UTA-VPRI 2023-037 REP/RES and UTA-VPRI 2024-920 IRP.
\bibliography{./references1}
\end{document}